\documentstyle[aps,epsf,pre,multicol]{revtex}
\draft

\begin{document}
\title{Rayleigh-B\'{e}nard convection in a homeotropically aligned nematic liquid crystal}
\author{Leif Thomas$^1$, Werner Pesch$^2$, and Guenter Ahlers$^1$}
\address{$^1$Department of Physics and Center for Nonlinear Science, University of California, Santa Barbara, California 93106}
\address{$^2$Institut f\"ur Theoretische Physik, Universit\"at Bayreuth, Bayreuth, Germany}
\date{\today}

\maketitle

\begin{abstract}

\noindent

We report experimental results for convection near onset in a thin
layer of a homeotropically aligned nematic liquid crystal heated from
below as a function of the temperature difference $\Delta T$ and the
applied vertical magnetic field $H$. When possible, these results are
compared with theoretical calculations. The experiments were done with three  cylindrical cells of
aspect ratios (radius/height) $\Gamma=10.6$, 6.2, and 5.0 over the
field range $8 \alt h \equiv H/ H_{F} \alt 80$ ($H_F =$ 20.9, 12.6,
and 9.3 Gauss are the Fr\'eedericksz fields for the three cells). We used the Nusselt number ${\cal N}$ (effective thermal
conductivity) to determine the critical Rayleigh number $R_c$ and the nature of the transition. We analysed digital images of the flow patterns to study the dynamics and to determine the mean wavenumbers of the convecting states. For $h$
less than a codimension-two field $h_{ct} \simeq 46$ the bifurcation
is subcritical and oscillatory, with travelling- and standing-wave transients. Beyond $h_{ct}$ the
bifurcation is stationary and subcritical until a tricritical field $h_t= 57.2$ is reached, beyond which it is supercritical. We analyzed the patterns to obtain the critical wavenumber $k_c$ and, for $h < h_{ct}$, the Hopf frequency $\omega_c$.In the subcritical range we used the early transients towards the finite-amplitude state for this purpose.  
The bifurcation sequence as a function of $h$ found in the experiment confirms the qualitative aspects of the theoretical predictions. Even
quantitatively the measurements of $R_c$, $k_c$ and $\Omega_c$ are
reproduced surprisingly well considering the complexity of the system.
However, the value of $h_{ct}$ is about 10\% higher than the
predicted value and the results for $k_c$ are systematically below the
theory by about 2\% at small $h$ and by as much as 7\% near
$h_{ct}$. At $h_{ct}$, $k_c$ is continuous within the experimental
resolution whereas the theory indicates a 7\% discontinuity. The
theoretical tricritical field $h_t^{th} = 51$ is somewhat below the
experimental one.
 The fully developed flow above $R_c$ for $h <
h_{ct}$ has a very slow chaotic time dependence which is unrelated to
the Hopf frequency.  For $h_{ct} < h < h_t$ the subcritical stationary
bifurcation also leads to a chaotic state.  The chaotic states persist
upon reducing the Rayleigh number below $R_c$, i.e. the bifurcation is hysteretic. 
Above the tricritical field $h_t$, we find a bifurcation to a time
independent pattern which within our resolution is
non-hysteretic. However, in this field range, there is a secondary
hysteretic bifurcation which again leads to a chaotic state observable
even slightly below $R_c$. We discuss the behavior of the system in the
high-field limit, and show that at the largest experimental field
values $R_c$ and $k_c$ are within 6\% and 1\%, respectively, of their
values for an infinite field.  

\end{abstract} \vskip 0.3in


\begin{multicols}{2}

\section{Introduction}
\label{sec:intro}

Convection in a thin horizontal layer of an isotropic fluid heated
from below by a heat current $\vec Q$ is well known as
Rayleigh-B\'{e}nard convection (RBC) \cite{Be00,RBC}.  When the fluid
is a nematic liquid crystal (NLC), this phenomenon is altered in
interesting ways.\cite{Pattern_Rev} NLC molecules are long, rod-like objects which are
orientationally ordered relative to their neighbors, but whose centers
of mass have no positional order \cite{Ge73,Bl83}. The axis parallel
to the average orientation is called the director $\hat n$. By
confining the NLC between two properly treated parallel plates
\cite{Co82}, one can obtain a sample with uniform planar (parallel to
the surfaces, i.e. in the x-y plane) or homeotropic (perpendicular to
the surfaces, or parallel to the z-axis) alignment of $\hat n$. The
alignment can be re-inforced by the application of a magnetic field
$\vec H$ parallel to the intended direction of $\hat n$. This is so
because the diamagnetic susceptibility is anisotropic, usually being
larger in the direction parallel to the long axis of the
molecules. The phenomena which occur near the onset of convection
depend on the orientation of $\hat n$ and $\vec
H$. \cite{Pattern_Rev} In this paper we are concerned with a
horizontal layer of a {\it homeotropically} aligned NLC in a vertical
magnetic field ($\vec H = H \hat e_z$) and heated from below. In that
case, $\vec Q = Q \hat e_z$ is parallel to $\hat n$ when the system is
in the conduction state. At a critical temperature difference $\Delta
T_c(H)$ the fluid will undergo a transition from conduction to
convection. The precise value of $\Delta T_c(H)$, the nature of the
bifurcation at $\Delta T_c(H)$, and the pattern-formation phenomena
beyond $\Delta T_c(H)$ are expected to depend in interesting ways upon
$H$ \cite{Le77,Le79,GPS79,FDPK92}.

A feature common to the homeotropic NLC and to an isotropic fluid
heated from below is that the system is isotropic in the horizontal
plane. Thus the convection pattern may form with no preference being
given to a particular horizontal axis unless the experimental
apparatus introduces an asymmetry. In both cases, the convection is
driven by the buoyancy force. However, the mechanism in the NLC case
is more involved.\cite{Le77,Le79,GPS79} The usual destabilization due
to the thermally-induced density gradient is opposed by the stiffness
of the director field which is coupled to and distorted by any
flow. Since relaxation times of the director field are much longer
than thermal relaxation times, it is possible for director
fluctuations and temperature/velocity fluctuations to be out of phase
as they grow in amplitude. The existence of two very different time
scales and this phase shift typically lead to an oscillatory
instability (also known as overstability), {\it i.e.} the bifurcation
at which these time-periodic perturbations acquire a positive growth
rate is a Hopf bifurcation.\cite{Le77,Le79,Hopf} This case is closely
analogous to convection in binary-fluid mixtures with a negative
separation ratio.\cite{HJ71,CH92} In that case, concentration
gradients oppose convection, and concentration diffusion has the slow
and heat diffusion the fast time scale. As in the binary mixtures, the
Hopf bifurcation in the NLC case is subcritical
\cite{GPS79,FDPK92}. For the NLC the fully developed nonlinear state
no longer is time periodic. Instead, the statistically stationary
state above the bifurcation is one of spatio-temporal chaos with a
typical time scale which is about two orders of magnitude slower than
the theoretical inverse Hopf frequency.\cite{Ah96} However, it is possible to
actually measure the Hopf frequency by looking at the growth or decay
of small perturbations which are either deliberately introduced
\cite{GPS79} or which occur spontaneously when the system is close to
the conduction state and near the bifurcation point.

A linear stability analysis of this system was carried out by several
investigators \cite{Le79,VZ79,BS81,BM83}. A very detailed analysis was
presented by Feng, Decker, Pesch, and Kramer (FDPK)
\cite{FDPK92}. These authors also provided a weakly-nonlinear
analysis, which allowed the distinction between sub- and supercritical
bifurcations.  Quantitative bifurcation diagrams were predicted for the
nematic liquid crystal MBBA.  In the present work we repeated and slightly extended the calculations for the material 5CB (see below) used in our experiments.  Since the material parameters of
MBBA and 5CB are similar, we found qualitatively the same bifurcation sequences as a function of the field.
Here we outline briefly the theoretical results and their relationship
to our experimental results.

As the magnetic field is increased, a subcritical Hopf
bifurcation-line is expected to terminate at $H_{ct}$ in a
codimension-two point (CTP) beyond which the perturbations which first
acquire a positive growth rate are at zero frequency.
Close to but beyond the CTP this stationary bifurcation is predicted
to also be subcritical. At an even higher field $H_t$ a tricritical
point (TCP) is predicted to exist  beyond which the
primary bifurcation is expected to become  supercritical. To our knowledge
there are no predictions about the patterns which should form beyond
either the subcritical Hopf bifurcation below the CTP or the
subcritical stationary bifurcation between the CTP and the
TCP. Although there are no explicit predictions of the patterns for $H
> H_{t}$, in analogy to isotropic fluids one might expect convection
rolls above the supercritical bifurcation, unless non-Boussinesq
effects\cite{OB,Bu67} yield a transcritical bifurcation to hexagons.

The phenomena described above were previously explored only partially
by experiment. Except for recent measurements at relatively small
fields \cite{Ah96}, the experiments have been qualitative or
semi-quantitative \cite{GPS79,SF97}. In the present paper we report
the results of an extensive sytematic experimental investigation of
this system which covered a wide range of magnetic fields $H$. In
agreement with previous work \cite{GPS79,SF97}, we find a subcritical
Hopf bifurcation at relatively small $H$ which terminates in a
codimension-two point (CTP). The CTP is located at a slightly higher
field than the theoretical prediction. We measured the Hopf frequency
$\omega_c(H)$ from visualizations of the spontaneous small-amplitude
early transients just above $\Delta T_c$. Except for the influence of
the small shift of the CTP, we found $\omega_c(H)$ to be in
quantitative agreement with the theory. From these transients, we also
determined the critical wavevector $k_c(H)$, and found it to be
typically a few percent smaller than the theoretical value. The reason
for this small difference between theory and experiment is not
known. As reported previously \cite{Ah96}, we found the convecting
nonlinear state for $\Delta T$ above $ \Delta T_c$ to be one of
spatio-temporal chaos (STC). Except at very small $H$, its
characteristic wavenumber was smaller than $k_c$ and insensitive to
$H$ and $\Delta T$. A long-time average of the structure factor of
this state was consistent with the expected rotational invariance of
the system. Depending on $H$, the lower limit $\Delta T_s$ at which
this chaotic state made its hysteretic transition back to the
conduction state was found to be 10 to 25\% below $R_c(H)$. The
convective heat transport was consistent with that of an isotropic
fluid with an average conductivity given by $\lambda_{avg} = (2
\lambda_\perp + \lambda_\parallel)/3$ where $\lambda_\perp$ and
$\lambda_\parallel$ are the conductivities perpendicular and parallel
to $\hat n$ respectively. This suggests a thorough randomization
of the director orientations by the flow.

Beyond the CTP we found a subcritical stationary bifurcation, as had
been predicted \cite{FDPK92}. The finite-amplitude state which evolved
was also a state of STC, but beyond a certain field value greater than
$H_{ct}$ it had distinctly different properties from the chaotic state
at lower $H$. This difference was clearly evident from a discontinuity
(as a function of $H$) of the characteristic wavenumber of the
nonlinear state, which was larger at the larger fields. There is no
theoretical guidance for the interpretation of these experimentally
observed phenomena.

The range $H \ge H_t$ was investigated for two cells with aspect ratios $\Gamma = 6.15$ and 5.01. We
refer to them as cells 5 and 6 respectively (for details see Sect.~\ref{sec:expt}
below). We found a primary bifurcation to a state with a
{\it hexagonal} flow pattern. Within our resolution this bifurcation
was non-hysteretic and the Nusselt number $\cal N$ grew continuously from zero.  The
appearance of hexagons rather than rolls is attributable to non-Boussinesq effects \cite{OB,Bu67} which occur when the up-down symmetry is
lifted by variations of the fluid properties over the cell height. Theoretically the bifurcation to hexagons is transcritical, and there should also be hysteresis associated with it. However, the hysteresis is often so small that it is unobservable even
though hexagons are found over a substantial range.\cite{Boden} At sufficiently large $\Delta T/\Delta T_c$ the hexagons become unstable with respect to rolls.\cite{Bu67} Since
 $\Delta T_c$ decreases with the cell thickness $d$ ($\sim d^{-3}$),
the range of stability of hexagons should depend on
the thickness of the fluid sample. However, for the {\it thinner} fluid layer
of cell 5, the existence range of hexagons was limited by a different secondary instability and grew from zero very near the TCP to $\epsilon \equiv \Delta T/\Delta T_c - 1 \simeq 0.1$ at the highest fields
available to us. At this stability limit a hysteretic transition yielded the chaotic state, and stationary rolls were never
found. With decreasing $\Delta T$, the chaotic state persisted down to
$\Delta T_s$ somewhat smaller than $\Delta T_c$. For the thicker fluid
layer of cell 6, typically $\Delta T_c$ was about 2 $^{o}$C, and
hexagons were found only up to $\epsilon \simeq 0.015$ even at high
fields. For $\epsilon > 0.015$, a pattern of rolls {\it not}
exhibiting STC was observed, as expected for a weakly non-Boussinesq system. At even higher $\epsilon$ and consistent
with the measurements in cell 5, a hysteretic secondary bifurcation
again yielded the chaotic state.

\section{Parameter Definitions and Values}                                   
\label{sec:paras}

The quantitative aspects of the instabilities are determined by four
dimensionless parameters which are formed from combinations of the
fluid properties \cite{FN1}. They are \cite{FDPK92} the Prandtl
number

\begin{equation} \sigma = {(\alpha_4/2)\over{\rho \kappa_\parallel}}\
\ , \label{eq:sigma} 
\end{equation}

\noindent the ratio between the director-relaxation time and the
heat-diffusion time

\begin{equation}
F = {(\alpha_4/2)\kappa _\parallel\over {k_{33}}}\ \ ,
\label{eq:F}
\end{equation}

\noindent the Rayleigh number

\begin{equation}
R = {\alpha g \rho d^3 \Delta T \over {(\alpha_4 /2)\kappa _\parallel}}\ ,
\label{eq:rayleigh}
\end{equation}

\noindent and the dimensionless magnetic field

\begin{equation}
h = H/H_{F}\ \ \ 
\label{eq:h}
\end{equation}

\noindent with the Fr\'eedericksz field

\begin{equation}
H_{F} = {\pi\over{d}}\sqrt{k_{33}\over{\rho \chi_a}}\ \ .
\label{eq:H_F}
\end{equation}

\noindent In these equations $\alpha_4$ is one of the viscosity
coefficients, $\kappa_\parallel$ is the thermal diffusivity parallel
to $\hat n$, $k_{33}$ is one of the elastic constants of the director
field, $\chi_a$ is the anisotropy of the diamagnetic susceptibility,
$\alpha$ is the isobaric thermal expension coefficient, and $g$ is the
gravitational acceleration.  The time scale of transients and pattern
dynamics is measured in terms of the thermal diffusion time

\begin{equation}
t_v = d^2/\kappa_\parallel\ \ .
\end{equation}

\noindent Both $h$ and $R$ are easily varied in an experiment, and may
be regarded as two independent control parameters. The availability of
$h$ in addition to $R$ makes it possible to explore an entire line of
instabilities.  The parameters $F$, $\sigma$, and $t_v$ are
essentially fixed once a particular NLC and temperature range have
been chosen, and even between different NLCs there is not a great
range at our disposal. For 5CB at 25.6$^\circ$ (the material and mean
temperature used in this work), we have $\sigma = 263$ and $F =
461$. The value of $t_v$ is typically several minutes, but depends on the thickness of the fluid layer. It is given in the next Section for each of our cells. The critical value $R_c(h)$ of $R$ and the fluid parameters
determine the critical temperature difference $\Delta T_c$ for a
sample of a given thickness $d$. The realistic experimental
requirement that $\Delta T_c \simeq $ a few $^\circ$C dictates that
the sample thickness should be a few $mm$.  Typical values of
$H_{F}$ are 10 to $20~Gauss$. Thus modest fields of a
$kGauss$ or so are adequate to explore the entire range of interest.

In order to evaluate $R_c$ from $\Delta T_c$, $h=H/H_F$,
and the theoretical values for $R_c(h), k_c(h),$ and $\omega_c(h)$, we
used the material properties given in Ref. \cite{FN1}. We
followed closely the calculational methods of FDPK. In order to insure
a sufficient resolution of any boundary layers we used Chebycheff
modes in the Galerkin method (no more than 20 were required).

\section{Experimental Apparatus and Sample Preparation}                                   
\label{sec:expt}

The apparatus used in this work was described previously
\cite{ACBS93,Ah96}. We made measurements using three circular cells of
different thicknesses, identified as cells 4, 5, and 6.
\cite{FN}  The thickness and radius were $d = 3.94\ mm$, $r = 41.9\
mm$ for cell 4, $d = 6.60\ mm$, $r = 40.6\ mm$ for cell 5, and $d =
8.88\ mm$, $r = 44.5\ mm$ for cell 6. The corresponding radial aspect
ratios $\Gamma \equiv r/d$ were 10.6, 6.15, and 5.01. The fluid was
4--n--pentyl--4$^\prime$--cyanobiphenyl (5CB).  All experiments were
performed at a mean temperature of 25.6$^\circ$C. The vertical thermal diffusion time was $t_v =$
139, 383, and 694 s, and $H_{F}$ was 20.1, 12.6, and 9.34 Gauss for cells 4, 5, and 6 respectively. Despite the longer time
scales involved for experiments in the thicker cells, cells 5 and 6 had an advantage over cell 4 due to the
smaller field strengths and temperature differences required
to perform the measurements. To insure homeotropic alignment near the
surfaces of the top and bottom plates of both cells, a surface
treatment with lecithin \cite{Co82,Ah96} was applied.

Defect-free homeotropic samples were prepared by applying a magnetic field 
 while cooling the bath, and thus the sapphire top plate of the
sample, from above the isotropic-nematic transition temperature
$T_{NI}$ to $T < T_{NI}$. During this process, the bottom plate
naturally lagged behind, and thus an adverse density gradient
existed. In the nematic/isotropic two-phase region even the relatively
small thermal gradients associated with small cooling rates induced
convection \cite{ABC93}. When the cooling was too rapid and the field
too small this led to a nematic sample with defects which remained
frozen.  By using cooling rates of $1^\circ$C/hour in the presence
of a field of $h \ge 17$ over the temperature interval 36 to
34$^\circ$C ($T_{NI} = 35.1^\circ$C) and annealing at 34$^\circ$C for
an hour or two the defects healed and a defect-free homeotropic sample
could be prepared.  Further cooling could then be at least ten times
as rapid without introducing new defects because the threshold for
convection in the nematic phase is large. Before each experimental
run, the procedure was repeated.

The critical temperature differences for the onset of convection were
determined from heat-transport measurements. These are usually
expressed in terms of the Nusselt number

\begin{equation}
{\cal N} \equiv \lambda_{eff} / \lambda_\parallel
\label{eq:nuss}
\end{equation}

\noindent where $\lambda_\parallel$ is the conductivity of the homeotropically aligned sample \cite{ACBS93}, and

\begin{equation}
\lambda_{eff} \equiv -Q d / \Delta T
\label{eq:lam_eff}
\end{equation}

\noindent is the effective conductivity and contains contributions
from diffusive heat conduction and from hydrodynamic-flow advection. Measurements of
$\cal N$ were made by determining the heat current $Q$ required to
hold $\Delta T$ constant. At each $\Delta T$, the heat current and
temperature of the bath and bottom plate were measured at one-minute
intervals for three to five hours, when typically all transients had died out.

In addition to heat-flow measurements, we also visualized the
convective flow patterns. The homeotropic samples were translucent
even for $d$ as large as several mm. It was just about possible to see
features of the bottom plate in typical ambient lighting. Any director distortion by convection rolls or domain walls
generated opaque regions with enhanced diffuse scattering which were easily visible. It should be kept in mind that
the optical signal in the images has a complicated relationship to the
hydrodynamic flow fields, and that quantitative information about
velocity- or temperature-field amplitudes could not be obtained. Such
quantities as the wavevector of the patterns or frequencies
of traveling convection-rolls could of course be determined
quantitatively.

The samples were illuminated from above by a circular fluorescent
light. Digital images were taken from above by
a video camera which was interfaced to a computer. Typically 50 to 200
images were averaged to improve the signal-to-noise ratio. Averaged
images were divided by an appropriate reference image to reduce the
influence of lateral variations in illumination and of other optical
imperfections. Some images were processed further by filtering in
Fourier space.

\narrowtext
\begin{figure}
\epsfxsize = 2.75in
\centerline{\epsffile{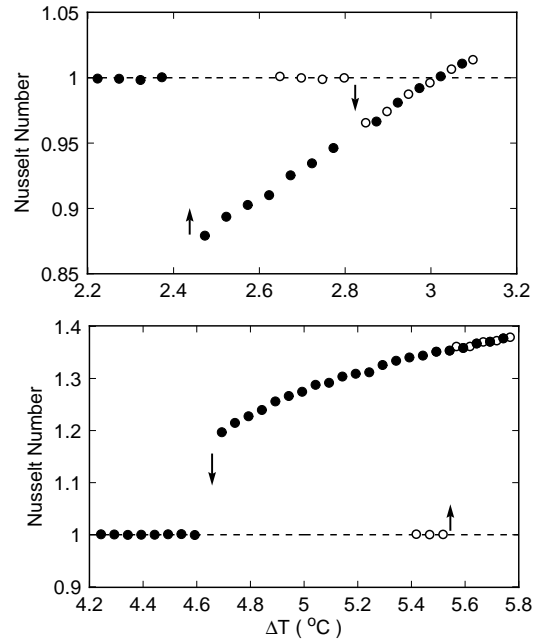}}  
\vskip 0.2in
\caption{Examples of the Nusselt number as a function of $\Delta T$ for cell 5. The upper figure is for $h = 15$ and the lower one is for $h = 50$. Open circles were taken with increasing and solid circles with decreasing $\Delta T$. The transitions between conduction and convection are indicated by the arrows.}
\label{fig:Nusselt1}
\end{figure}

\section{Results}
\label{sec:results}

\subsection{Nusselt Numbers and Critical Rayleigh Numbers}

Figure~\ref{fig:Nusselt1} shows $\cal N$ for cell 5 as a function of
$\Delta T$ for two field strengths $h = 15$ and $h = 50$. The open
circles were obtained with increasing, and the solid circles with
decreasing $\Delta T$. For the lower fields ($h < 20$) $\cal N$
decreased below one when convection started. This can be understood
because the convecting sample has a distorted director with a component perpendicular to $\vec Q$. The contribution from this component to the conductivity corresponds to $\lambda _{\perp}$, which is less
than the conductivity $\lambda _\parallel$ of the homeotropic case.
\cite{ACBS93} It turns out that for small fields the direct hydrodynamic contribution to
the heat flux is smaller than the decrease in the heat flux due to the
director distortion by the
flow.  For the higher fields ($h > 35$), $\cal N$ remained above one in
the convecting state. Thus with the higher fields the hydrodynamic
contribution to the heat flux is greater than the decrease in the heat
flux due to any distortion of the director.  Both examples in
Fig.~\ref{fig:Nusselt1} demonstrate the predicted and previously
observed \cite{GPS79,Ah96,SF97} hysteretic nature of the bifurcation, i.e. as
$\Delta T$ was decreased, the conduction state was reached at a value
of $\Delta T$ equal to $\Delta T_s < \Delta T_c$.

\begin{figure}
\epsfxsize = 3in
\centerline{\epsffile{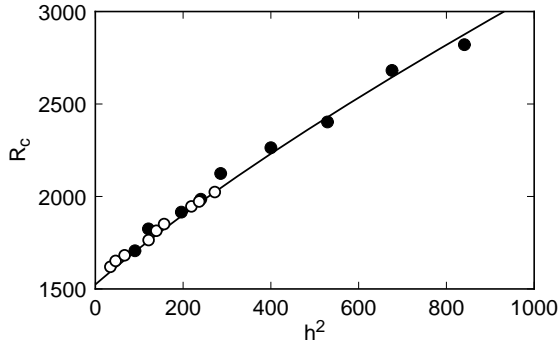}}  
\vskip 0.2in
\caption{Critical Rayleigh numbers for the onset of convection as a function of $h^2$. Open and filled symbols were obtained in cells 4 and 5, respectively. The line is the theoretical prediction.}
\label{fig:R_c_v_h^2}
\end{figure}

From data like those in Fig.~\ref{fig:Nusselt1}, critical temperature
differences $\Delta T_c$ were determined with an uncertainty of less
than one percent. The corresponding Rayleigh numbers are shown in
Fig.~\ref{fig:R_c_v_h^2} as a function of $h^2$. The open circles were
obtained in cell 4, the filled ones in cell 5. The good agreement
between the two data sets confirms the expected scaling of the field
with $H_F$. It also shows that using the fluid properties at the mean
temperature does not lead to systematic errors in $R_c$ even for cell
4 where $\Delta T_c$ is over 10$^\circ$C. One sees that $R_c$ is
quadratic in $h$ at small $h$, as is expected because the system
should be invariant under a change of the field direction. The solid
line follows the theoretical prediction and  agreement between
theory and experiment is excellent.

\begin{figure}
\epsfxsize = 3in
\centerline{\epsffile{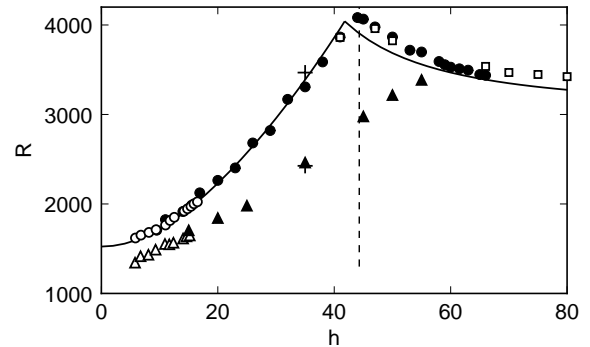}}  
\vskip 0.2in
\caption{Critical Rayleigh numbers $R_c$ and saddle-node Rayleigh numbers $R_s$ over the experimentally accessible field range. The open circles, filled circles, and open squares are $R_c$ for cells 4, 5, and 6 respectively. The open and filled triangles are $R_s$ for cells 4 and 5 respectively. The dashed line indicates the location of the codimension-two point as found experimentally. The solid line is the theoretical prediction for $R_c$. The plusses at $h = 35$ were obtained with short equilibration times and cell 5 (see text).}
\label{fig:Bif_dia}
\end{figure}

Results for $R_c$ over our full experimental field range are shown as
a function of $h$ in Fig.~\ref{fig:Bif_dia}. Here we include data
taken with cell 6 as open squares. The data for $R_c$ reveal a sharp
maximum at $h = 44.3$. We interpret this field value as the
codimension-two point $h_{ct}$ and indicate it in
Fig.~\ref{fig:Bif_dia} by the dashed vertical line. The solid line in
the figure is the theoretical prediction for $R_c$, evaluated for the
properties of our sample. For the entire range $h < h_{ct}$, the
theoretical result is in excellent agreement with the data. However,
the theory gives $h_{ct} = 41.8$ which is slightly lower than the
experimental value. Above $h_{ct}$ the measurements of $R_c$ are
systematically larger than the calculation, although the largest
discrepancy is only about 4\%.

The triangles in Fig.~\ref{fig:Bif_dia} show the lower limit of
existence (the ``saddle-node" Rayleigh number $R_s$) of the finite-amplitude convecting state as determined from
data like those in Fig.~\ref{fig:Nusselt1}. They suggest that the
tricritical bifurcation is located near $h = 59$, which is larger than
the theoretically calculated value $h_t \simeq 51$. However, we will
return later to the best estimate of $h_t$.

\begin{figure}
\epsfxsize = 3in
\centerline{\epsffile{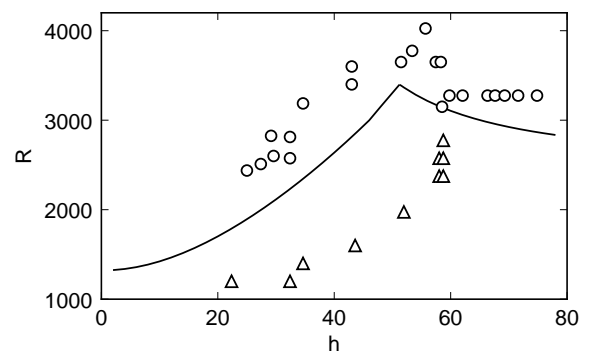}}  
\vskip 0.2in
\caption{Results for $R_c$ and $R_s$ from Ref. \protect{\cite{SF97}} obtained with MBBA. The theoretical curve for $R_c$ was computed from typical fluid properties of MBBA \protect{\cite{FN2}}.}
\label{fig:Compare}
\end{figure}

Measurements similar to those shown in Fig.~\ref{fig:Bif_dia} were
made by Sal\'an and Fern\'andez-Vela \cite{SF97} (SF), using the
nematic liquid crystal MBBA. Their results are shown in
Fig.~\ref{fig:Compare}, together with the theoretical curve for that
case \cite{FN2}. The data and the curve illustrate that there are
significant quantitative differences between the bifurcation lines of
different nematics. In Fig.~\ref{fig:Compare} the experimental points
lie on average about 25\% above the theoretical curve, and the lower
hysteresis limit is further below the bifurcation line than we found
for 5CB. 

\noindent The equilibration times after each temperature step used by
SF were 30 minutes, which is a factor of six to ten shorter than
those of our experiments. In addition the temperature steps of SF were
a factor of two larger than ours, yielding a difference in the average
rate of change of the temperature of a factor of 12 or more. Looking for
an explanation of the difference between the experimental and
theoretical $R_c$ revealed in Fig.~\ref{fig:Compare}, we conducted one
run with equilibration times similar to those of SF, but using our 5CB
sample. It gave the plusses in Fig.~\ref{fig:Bif_dia}. As can be seen,
these results do not differ significantly from the data taken with our
usual longer equilibration times. Thus we have no explanation for the
difference between the SF data for $R_c$ and the theoretical
curve. However, the agreement between our runs with the
different equilibration times implies that our usual experimental
procedure yielded quasi-static results.

\begin{figure}
\epsfxsize = 3in
\centerline{\epsffile{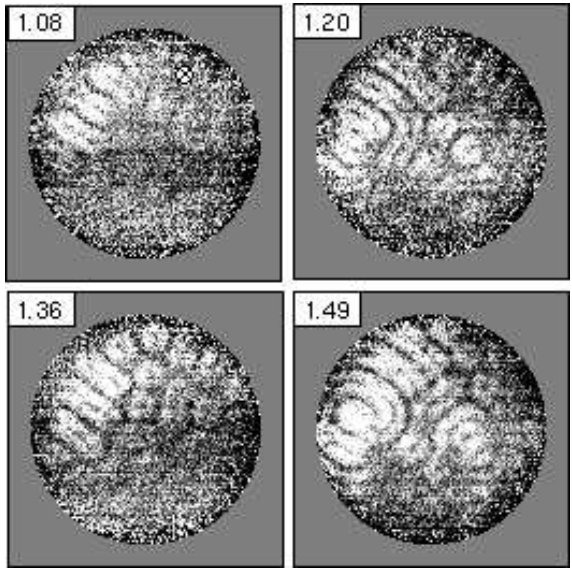}}  
\vskip 0.2in
\caption{A sequence of images of the travelling/standing wave transients for $h = 32$ and $\epsilon = 0.015$ in cell 5. The number in each image corresponds to the time, in units of $t_v$, that elapsed since flow first became visible. A time series of the pixel intensity was taken at the location marked in the top left image.}
\label{fig:Transients1}
\end{figure}

\subsection{Hopf Frequency and Critical Wavevector}

\begin{figure}
\epsfxsize = 3in
\centerline{\epsffile{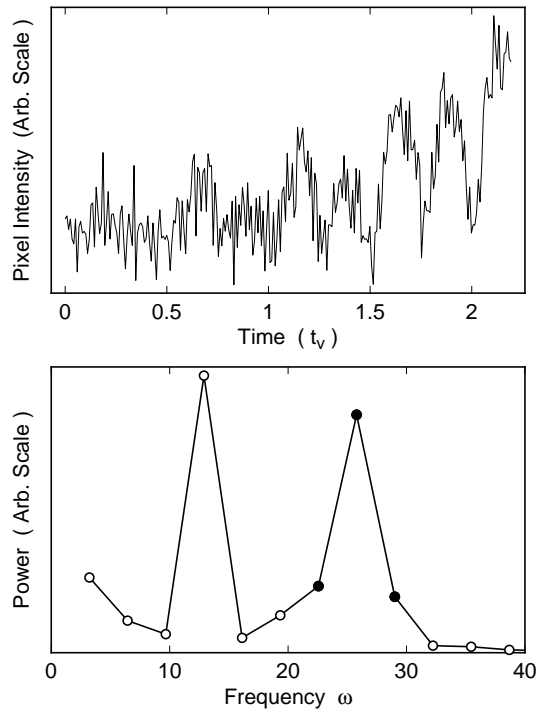}}  
\vskip 0.2in
\caption{The upper figure is the time series of the pixel intensity at the location shown in the top left image of Fig.\protect{\ref{fig:Transients1}}. The lower figure is the power spectrum of that time series. The mean frequency calculated from the solid circles was used to determine the Hopf frequency. }
\label{fig:Frequency}
\end{figure}

Once the critical Rayleigh numbers were measured, a detailed analysis
of the patterns could be undertaken. At first we will
characterize the Hopf bifurcation for $h$ below
$h_{ct}$. Since in that field range the bifurcation is subcritical,
we had to use the small-amplitude transients to determine $\omega_c$ and $k_c$. 
Figure~\ref{fig:Transients1} shows images from cell 5 which are
characteristic of these patterns. They were taken at the times (in
units of $t_v = 383$~s) indicated in each figure after the pattern initially
became visible.  This typically occurred around one hour after
$\epsilon$ was raised from below zero to around 0.015. Inspection of
successive images revealed that the transients could be either
traveling or standing waves, sometimes with both occurring at
different locations in the same cell.  In the top left image of
Fig.~\ref{fig:Transients1}, a location is indicated at which a time
series of the pixel intensity was acquired. This time series is shown
in Fig.~\ref{fig:Frequency} along with its corresponding power
spectrum.  The length of the time series was limited by the rapid
growth of the pattern to its finite-amplitude steady state.  Because
of this, only a small number of periods could be obtained before the
finite-amplitude state was reached. Thus to avoid errors associated
with incommensurate sampling, the data was windowed before its Fourier
transform was evaluated. This process was repeated at several pixel
locations in the cell. The signal from the second harmonic was often
found to be stronger than that from the fundamental. Thus it was
used to calculate the frequency.  The frequencies at different
locations generally were within a percent of each other, and were
averaged to determine the critical Hopf frequency $\omega_c$.

\begin{figure}
\epsfxsize = 3in
\centerline{\epsffile{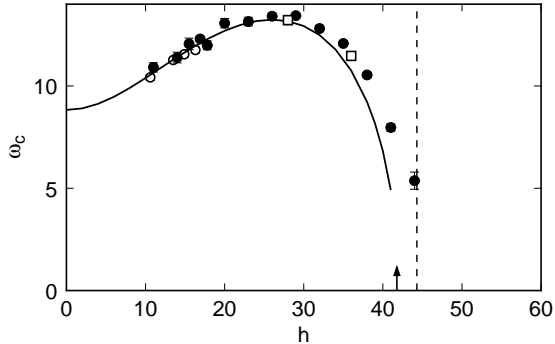}}  
\vskip 0.2in
\caption{The Hopf frequency $\omega_c$ as a function of $h$. Open and filled circles are for cells 4 and 5, respectively. Open squares are for cell 6. The solid line is the theoretical prediction for $\omega_c$. The dashed line (arrow) indicates the location of the codimension-two point as found experimentally (theoretically).}
\label{fig:Omega_c}
\end{figure}

The dependence upon $h$ of the measured $\omega_c$ is compared with
theory in Fig.~\ref{fig:Omega_c}.  The arrow indicates the location of
the theoretical codimension-two point while the dashed line represents
the experimental determination of $h_{ct}$. As can be seen, away from
the codimension-two point the agreement with the measurements is
excellent. In accordance with theory, the experimental $\omega_c$
changes discontinuously to zero at $h_{ct}$, above which the bifurcation
is stationary.

\begin{figure}
\epsfxsize = 3in
\centerline{\epsffile{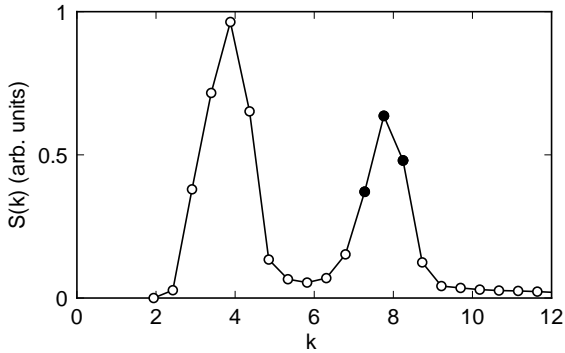}}  
\vskip 0.2in
\caption{The azimuthal average of the time-averaged structure factor $<S(k)>$ of the travelling/standing wave transients at $h = 32$ for cell 5. The mean wavenumber calculated from the solid circles was used as the critical wavenumber.}
\label{fig:Struct_factor}
\end{figure}

By evaluating the Fourier transforms of images such as those in
Fig.~\ref{fig:Transients1}, the critical wavenumber $k_c$ of the
patterns could be measured.  The transforms were based on the central
parts of the images by using the filter function $W(r) = \{1 +
cos[(\pi)(r/r_o)]\}/2$ for $r < r_o$ and $W(r) = 0$ for $r > r_o$.
Here $r_o$ was set equal to 85\% of the sample radius. Time averaging
the square of the modulus of the transforms over the length of the
time series yielded the structure factor $<S(\bf k)>$.
Figure~\ref{fig:Struct_factor} shows the azimuthal average $<S(k)>$ of
$<S(\bf k)>$ for the run at $h = 32$. We used a weighted average of
the three points nearest the peak of the second harmonic of $<S(k)>$
to calculate $k_c$. 

\begin{figure}
\epsfxsize = 3in
\centerline{\epsffile{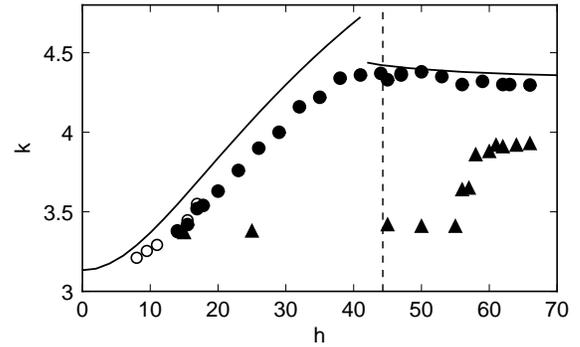}}  
\vskip 0.2in
\caption{The characteristic wavenumbers of the observed patterns as a function of $h$. Circles: The wavenumber of the small-amplitude transients. Triangles: The wavenumber $k_n$ of the fully developed spatially and temporally chaotic flow, as measured close to the onset of such flows. Open and filled symbols were obtained in cells 4 and 5, respectively.}
\label{fig:K_v_h}
\end{figure}

Figure~\ref{fig:K_v_h} displays the results for $k_c$ for all $h$
together with the theoretical analysis.  For $h < h_{ct}$ the measured
critical wavenumber of the transients is systematically smaller than
the theoretical one. When the codimension-two point is approached,
the experimental wavenumbers make a smooth rather than discontinuous
transition to those associated with the stationary bifurcation,
whereas the theory predicts a 7\% discontinuity of $k_c$ at $h_{ct}$.
The reason for these discrepancies is as yet unknown. Above $h_{ct}$,
the agreement between the experimental and theoretical wavenumbers is
excellent.

As shown explicitly for $R_c$ in Fig.~\ref{fig:R_c_v_h^2}, $R_c, \omega_c$, and $k_c$ are proportional to $h^2$ for small $h$.  

\subsection{Nonlinear State below the tricritical field $h_{ct}$}
  
\begin{figure}
\epsfxsize = 3in
\centerline{\epsffile{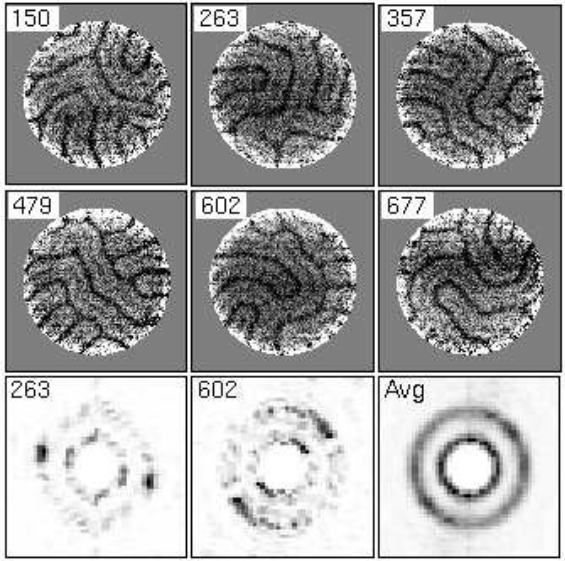}}  
\vskip 0.2in
\caption{Top two rows: a sequence of images taken with constant external conditions ($h = 50$, $\epsilon = 0.014$) for cell 5. The time elapsed since $\epsilon$ was raised from below zero (in units of $t_v = 383s$) is given in the top left corner of each image. Bottom row: The structure factor of two of the images shown above, and the average of the structure factor of 75 images spanning a time interval of 724$t_v$. The structure factor was obtained using a Hanning window, and thus is dominated by the patterns near the cell center.}
\label{fig:STC_sequence}
\end{figure}

  Because of the subcritical nature of the bifurcation for $h < h_{ct}$ a
  finite-amplitude state develops directly at onset.  The time
  dependence and spatial structure of this state are very different
  from that of the small-amplitude transient state. The first two rows
  of Fig.~\ref{fig:STC_sequence} show typical images of the patterns
  from cell 5 which are characteristic of the fully developed flow.
  They are from a single experimental run with constant external
  conditions. They were taken at the times indicated in each image, in
  units of $t_v = 383~s$, which had elapsed since $\epsilon$ had been
  raised from below zero to 0.014. The convection rolls of the fully
  developed flow show an irregular time dependence, with typical time
  scales around a hundred times longer than the inverse Hopf
  frequencies of the transients. One can see that the ``chaotic'' behavior is
  associated with the formation of defects and the continuous
  reorientation of the convection rolls. This continuous reorientation
  of the rolls is evident in the rightmost image in the bottom row of
  Fig.~\ref{fig:STC_sequence} (labeled ``Avg"). It shows the time
  average of the structure factor. The average involved 75 images
  taken over a total time period of $724t_v$ (over three days). It is
  seen to contain contributions at all angles, consistent with the
  idea of a statistically stationary process of non-periodic pattern
  evolution and with the expected rotational symmetry of the system.
  Similar results for cell 4 have been shown previously.\cite{Ah96}.

\begin{figure}
\epsfxsize = 3in
\centerline{\epsffile{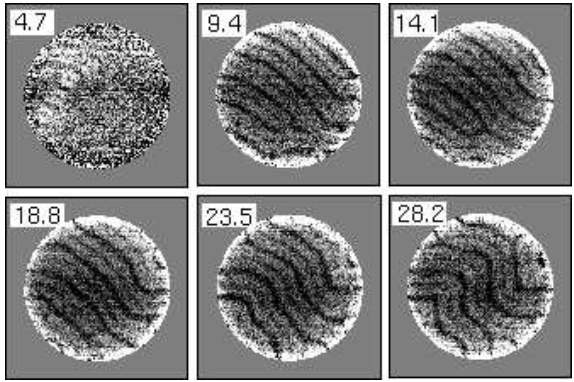}}  
\vskip 0.2in
\caption{A temporal succession of images during the transient leading from conduction to convection when $\Delta T$ was raised slightly above $\Delta T_c$ for cell 5. The field was $h = 50$. The numbers are the elapsed time, in units of $t_v$, since the threshold was exceeded.}
\label{fig:Transients2}
\end{figure}

When $h$ was increased above $h_{ct}$, the nature of the pattern at first did not change noticeably. For instance, as evident in
Fig.~\ref{fig:K_v_h}, the characteristic wavenumber of the pattern (as
denoted by the triangles) remained close to 3.4 for $h
\le 55$. Over this field range the patterns of the fully
developed flow look similar to those illustrated in
Fig.~\ref{fig:STC_sequence}, i.e. they exhibit spatio-temporal chaos.

\begin{figure}
\epsfxsize = 3in
\centerline{\epsffile{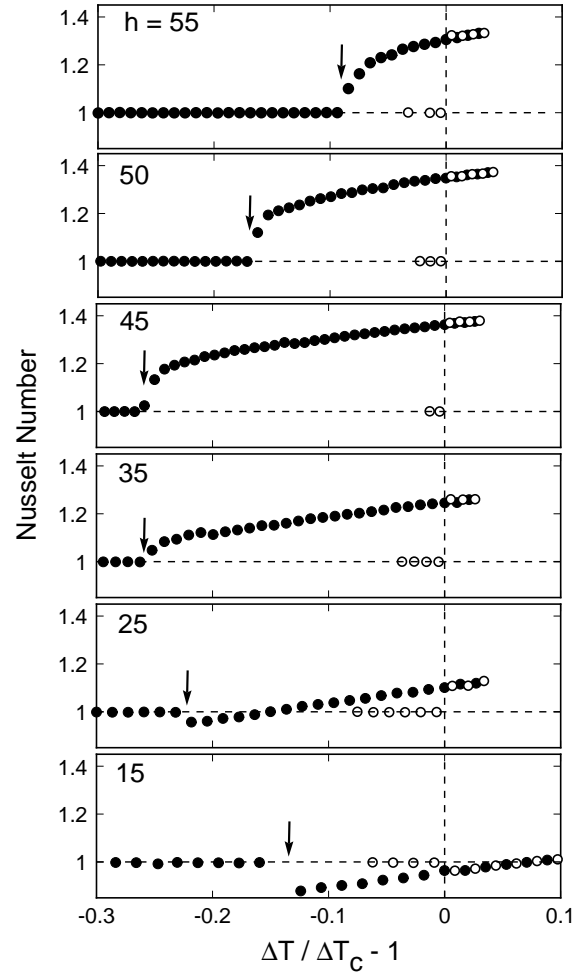}}  
\vskip 0.2in
\caption{Nusselt-number measurements for cell 5 illustrating the variation of the size of the hysteresis loop between conduction and convection with $h$. The number in the upper left corner of each plot is the field $h$. Open circles were taken with increasing and solid circles with decreasing $\Delta T$. The arrows show the values of $\epsilon_s \equiv R_s/R_c - 1$.}
\label{fig:Nusselt2}
\end{figure}

It is instructive to examine the transients which lead from the
small-amplitude to the finite-amplitude statistically-stationary
state. This is done in Fig.~\ref{fig:Transients2}. Here the number in
each image gives the time, in units of $t_v$, which had elapsed since
$\Delta T$ was raised slightly (1.4\%) above $\Delta T_c$. At $t =
4.7$ small-amplitude transients like those in Fig.~\ref{fig:Transients1} are evident in part of the cell.
By $t = 9.4$ these had filled the cell and grown to a saturated amplitude. At this stage they formed nearly-straight
parallel rolls with a wavenumber which was smaller than $k_c$. However, these straight
rolls turned out to be unstable to a zig-zag instability. In the end,
this instability led to the spatially and temporally disordered
pattern as shown in Fig.~\ref{fig:STC_sequence}. Thus, we see that a
secondary instability led to a chaotic state rather than to a new
time-independent pattern. This phenomenon most likely is similar to
the one encountered in very early experiments on spatio-temporal chaos
using liquid helium \cite{Ah74a,Ah74b}, where ordinary RB convection
became chaotically time dependent, most likely because the secondary 
skewed-varicose instability \cite{BC79a} was crossed.

\begin{figure}
\epsfxsize = 3in
\centerline{\epsffile{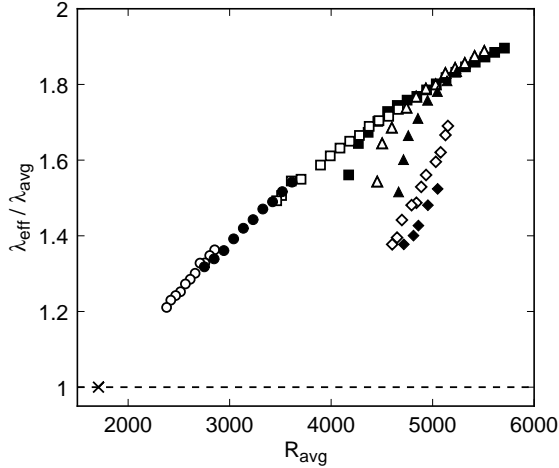}}  
\vskip 0.2in
\caption{Nusselt-number measurements for cell 5, normalized by the averaged conductivity $\lambda_{avg} \equiv (2\lambda_\perp + \lambda_{\parallel})/3$, as a function of the Rayleigh number $R_{avg}$ computed with $\kappa_{avg} \equiv \lambda_{avg}/\rho C_P$. The data are for h = 15 (open circles), 25 (solid circles), 35 (open squares), 45 (solid squares), 50 (open triangles), and 55 (solid triangles). The diamonds are for h = 64, which is in the supercritical region. Here the solid diamonds are for the hexagons/rolls which form supercritically, and the open ones are for the chaotic finite-amplitude state which forms via a secondary bifurcation (see Fig.~\protect{\ref{fig:Nusselt4}} below). The cross corresponds to the critical Rayleigh number $R_c = 1708$ of an isotropic fluid.}
\label{fig:Nusselt2b}
\end{figure}

Heat-transport measurements of the fully developed flow are shown in
Fig.~\ref{fig:Nusselt2} for several field values as a function of
$\epsilon$. They illustrate the evolution with $h$ of the hysteretic
nature of the bifurcation. As can be seen also in
Fig.~\ref{fig:Bif_dia}, the hysteresis $|\epsilon_s|$ increased with $h$ for $h < h_{ct}$ from about 10\% at the low fields to nearly 25\% close to the
codimension-two point.  Above this point the hysteresis decreased and
suggested the existence of a tricritical point near $h_{t} \simeq 59$ (see below for more detail about the tricritical region).

When the Nusselt numbers in Fig.~\ref{fig:Nusselt2} are plotted
against the Rayleigh number $R$, they fall on or approach a single
curve independent of $h$. This suggests that the convection in the
chaotic state is sufficiently vigorous to achieve nearly complete
randomization of the director orientations, regardless of $h$. In that
case one would expect that the system should behave approximately like
an isotropic fluid, with an averaged conductivity $\lambda_{avg} =
(2\lambda_\perp + \lambda_\parallel)/3$. Thus we plot in
Fig.~\ref{fig:Nusselt2b} a modified Nusselt number $\tilde {\cal N}$
given by the ratio of the effective conductivity of the convecting
state to $\lambda_{avg}$ as a function of $R_{avg}$, where $R_{avg}$
is computed using $\kappa_{avg} = \lambda_{avg}/\rho C_P$ in
Eq.~\ref{eq:rayleigh} rather than $\kappa_{\parallel}$. At all but the
highest fields (where the primary bifurcation is supercritical) the
data reach the common curve. At small $R_{avg}$, this curve
extrapolates to $\tilde {\cal N} = 1$ near $R_{avg} = 1708$ (the cross
in the figure), which is the critical Rayleigh number of an isotropic
fluid. An analogous behavior has been observed in binary-mixture
convection with negative separation ratios $\Psi$,\cite{Surko} where
the bifurcation is also subcritical. In that case the convective flow
achieves thorough mixing of the concentration field and $\cal N$
approaches a curve which is independent of $\Psi$. In both cases the
mixing achieved by the flow can persists because of the existence of a
slow time scale, namely that of director or concentration relaxation.

Further support for the idea that the chaotic flow in some respects can be approximated
by isotropic-fluid convection is found in Fig.~\ref{fig:K_v_h}, where
for $h \alt 55$ the wavevectors (triangles) are independent of $h$ and
much closer to the critical value $k_c^{iso} = 3.117$ than to the
critical values $k_c(h)$ of the anisotropic system (circles in
Fig.~\ref{fig:K_v_h}). Exact agreement with $k_c^{iso}$ would of
course not be expected even for a genuine isotropic fluid because of
the finite flow amplitude and various wavenumber-selection processes.

Lastly we note that an extrapolation of the data in
Fig.~\ref{fig:Nusselt2b} to $\tilde {\cal N} = 1$ and $R_{avg} = 1708$
yields an initial slope $\tilde S_1$ of $\tilde {\cal N}- 1 = \tilde S_1
(R_{avg}/1708 - 1)$ of about 0.6. For a laterally infinite system of
straight rolls in an isotropic fluid with a large Prandtl number one
expects $S_1 \simeq 1.43$.\cite{SLB65} However, experiments in finite
systems with modest aspect ratios\cite{MAC91} have always yielded
smaller values, usually in the range of 0.6 to 1. Particularly when
many defects are present, as in our case, one would expect the heat
transport to be suppressed relative to that of a perfect straight-roll
structure.

\subsection{Tricritical Region and beyond}

This section is devoted to the phenomena which occur near the 
tricritical field $h = h_{t}$. At first the Nusselt numbers and the
patterns are described and the corresponding
bifurcation diagram is given. Further subsections deal with the
precise determination of $h_{t}$ and with hexagons observed near
threshold for $h > h_{t}$.

\subsubsection{Nusselt Numbers and Patterns}

\begin{figure}
\epsfxsize = 3in
\centerline{\epsffile{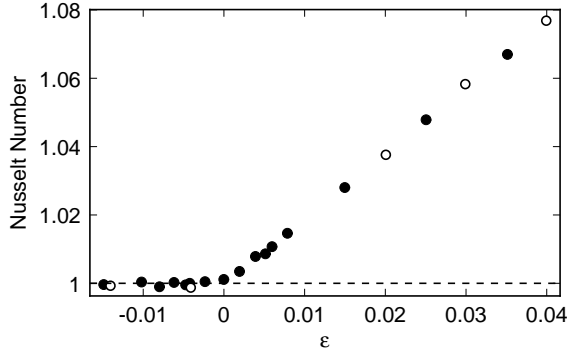}}  
\vskip 0.2in
\caption{Nusselt-number measurements for $h = 63$ and cell 5 illustrating the supercritical nature of the bifurcation characteristic of the high fields. Open circles were taken with increasing and solid circles with decreasing $\Delta T$.}
\label{fig:Supercrit}
\end{figure}

From the measurements of the Nusselt number (see
Fig.~\ref{fig:Nusselt2}) there is clear evidence of a tricritical
field $h_{t}$, above which the primary bifurcation is
supercritical. For instance for $h \simeq 60$, measurements of 
${\cal N}$ revealed no hysteresis at the primary
bifurcation and within our resolution ${\cal N}$ grew continuously from
one beyond $\Delta T_c$. This is exemplified for
cell 5 and $h = 63$ in Fig.~\ref{fig:Supercrit}. The open (solid) circles
correspond to the stable states reached by increasing (decreasing)
$\Delta T$.\cite{FN3}  This behavior of ${\cal N}$ stands in sharp contrast to
that shown in Fig.~\ref{fig:Nusselt2} for lower fields.
 
\begin{figure}
\epsfxsize = 3.5in
\centerline{\epsffile{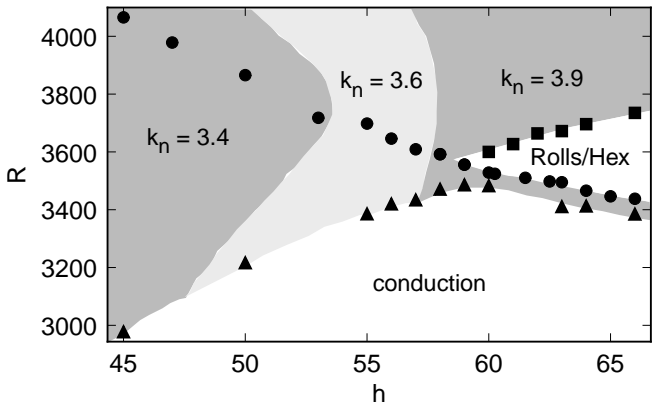}}  
\vskip 0.2in
\caption{Bifurcation diagram in the region of the $R - h$ plane close to the tricritical point. Solid circles: primary bifurcation. Solid triangles: $R_s$. Solid squares: hysteretic secondary bifurcation to chaotic convection. Shaded areas labeled $k_n = 3.4$, 3.6, and 3.9 correspond to chaotic regimes with different mean wavenumbers. The wedge-shaped area labeled Rolls/Hex shows the parameter range over which time-independent  convection is stable. For cell 5, the pattern is hexagonal in this entire region. For cell 6, the pattern is hexagonal in this region for $\epsilon \leq 0.015$. For larger $\epsilon$ but still in this region it consists of time-independent rolls.}
\label{fig:Bif_dia_detail}
\end{figure}

Besides the Nusselt number the analysis of the patterns gives
important additional insight in particular with respect to secondary
transitions.  In Fig.~\ref{fig:Bif_dia_detail} all the available information has
been condensed in a bifurcation diagram for the vicinity of the
tricritical point.   

\begin{figure}
\epsfxsize = 2in
\centerline{\epsffile{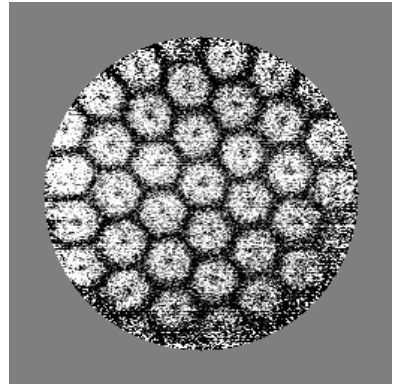}}  
\vskip 0.2in
\caption{Image of the hexagonal flow in cell 5 for $h = 65$ and $\epsilon = 0.01$. The pattern was essentially the same over the entire existence range of hexagons.}
\label{fig:Hexagons}
\end{figure}

\begin{figure}
\epsfxsize = 3in
\centerline{\epsffile{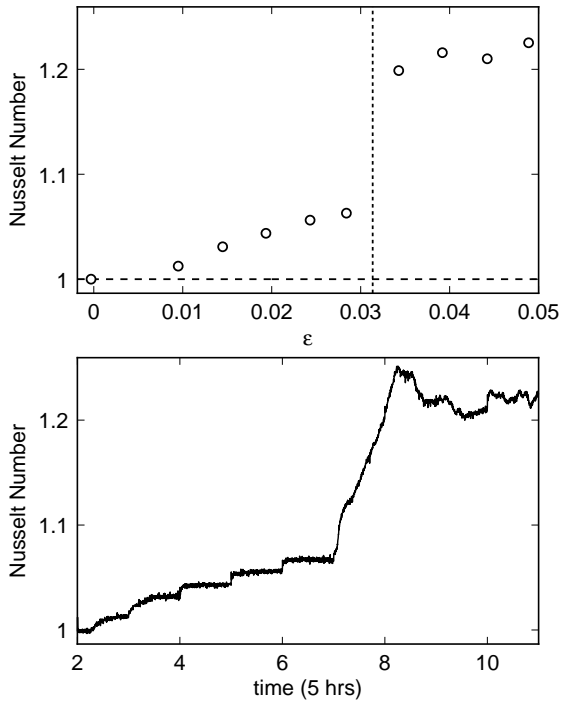}}  
\vskip 0.2in
\caption{Nusselt-number measurements for $h = 61$ in cell 5. The upper figure illustrates the dependence of the Nusselt number on $\epsilon$. The dotted line indicates $\epsilon_n$, where the transition from hexagons to rolls occurred when $\Delta T$ was increased. The lower figure illustrates the dependence of the Nusselt number on time. Time is measured in units of 5 hours, i.e. the time between steps in $\epsilon$. }
\label{fig:Nusselt3}
\end{figure}

\begin{figure}
\epsfxsize = 3in
\centerline{\epsffile{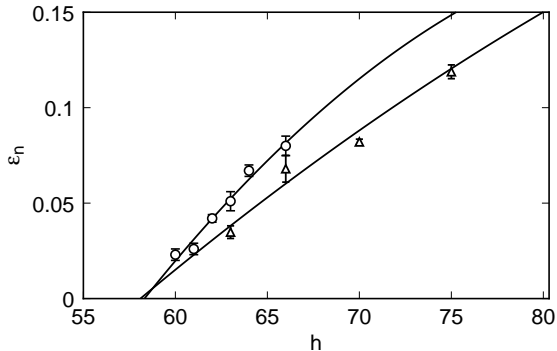}}  
\vskip 0.2in
\caption{Values of $\epsilon_n(h)$ where a transition to a spatially and temporally chaotic state occurred. Circles and triangles were obtained in cells 5 and 6, respectively. The lines represent a fit of a quadratic polynomial in $(h/h_n - 1)$ to the data. The fits extrapolate to zero at $h_n = 58.3 \pm 0.7$ for cell 5 and $h_n = 58 \pm 3$ for cell 6.}
\label{fig:Epsilon_n}
\end{figure}

At first we will focus on the wedge labeled Rolls/Hex where time-independent
convection is stable.  For both cell 5 and 6, a seemingly supercritical
primary bifurcation led to a hexagonal pattern. For cell 5
this pattern is shown in Fig.~\ref{fig:Hexagons}. The range of
$\epsilon$ over which the hexagons were stable differed in the two
cells. In cell 5, hexagons remained stable up to $R_n(h)$
(solid squares in Fig.~\ref{fig:Bif_dia_detail}) for the entire range
of $h$. At $R_n(h)$ a transition to a spatially and temporally
chaotic roll pattern with a lower characteristic wavenumber occurred.
For $h < 75$ this transition 
was distinguished by a jump in ${\cal N}$
as well as the onset of time dependence of ${\cal N}$, as illustrated
in Fig.~\ref{fig:Nusselt3}. The upper figure gives the steady-state
${\cal N}$ and shows the    
jump at $\epsilon_n \equiv R_n/R_c - 1$. The lower figure is the
time series of ${\cal N}$ obtained in the same run. Here $\epsilon$
was held constant for a five-hour period at each of the eleven
successively increasing values. The data show that ${\cal N}$ is
steady below and time dependent above $\epsilon_n(h)$. For $h \ge 75$
the discontinuity in ${\cal N}$ was no longer pronounced, but a
transition to time dependence still occurred at $\epsilon_n(h)$. Thus,
depending on the field, either of these two indicators was used to
determine the location of $\epsilon_n(h)$. In the thicker cell 6, a
transition from hexagons to rolls occurred near $\epsilon \simeq
0.015$, independent of field strength. This transition was not
associated with a measurable change in the wavenumber, a jump in
${\cal N}$, or a time dependence of ${\cal N}$. A further increase of
$\epsilon$ again led to a transition at $\epsilon_n(h)$ from steady rolls to
the chaotic state, consistent with the cell 6 experiments. The results
for $\epsilon_n(h)$ obtained in cells 5 and 6 are shown in
Fig.~\ref{fig:Epsilon_n} as circles and triangles respectively.

On the basis of the usual Landau equation for a tricritical
bifurcation, one would expect the hysteresis to grow gradually as $h$
is reduced below $h_t$. However, within our resolution this was not
the case and a hysteretic primary bifurcation to a chaotic state
occurred immediately below $h_t$. Indeed, the secondary bifurcation
line $\epsilon_n(h)$ for $h > h_t$ met the primary bifurcation line at
$h_t$ within experimental resolution, as can be seen already in
Fig.~\ref{fig:Bif_dia_detail}.

\noindent It is shown more explicitly in
figure~\ref{fig:Epsilon_n}, where the range $\epsilon_n$ of
time-independent patterns vanishes near $h = 58$. Fitting a quadratic polynomial 
in $(h/h_n -1 )$ to $\epsilon_n(h)$ yields $h_n = 58.3 \pm 0.7$ (cell
5) and $h_n = 58 \pm 3$ (cell 6) for the field where $\epsilon_n$
vanishes. Within error, these values agree with the tricritical field
obtained from the slope of the Nusselt number (see the next Section).The bifurcation for $h > h_t$ is supercritical but the amplitude at constant $\epsilon > 0$ diverges as $h$ approaches $h_t$ from above. Since secondary bifurcations occur at finite values of the amplitude, we expect $\epsilon_n(h)$ to vanish at $h_t$. 
Therefore the $\epsilon_n$ measurements provide a relatively precise
lower limit $h_t = 57.6$ for the tricritical field.

\begin{figure}
\epsfxsize = 3in
\centerline{\epsffile{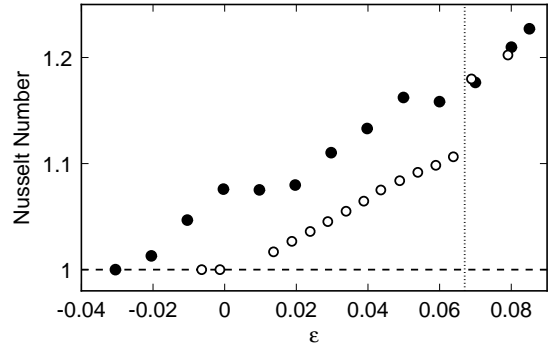}}  
\vskip 0.2in
\caption{Nusselt-number measurements for $h = 64$ in cell 5. Open circles were taken with increasing and solid circles with decreasing $\Delta T$. The dotted line indicates $\epsilon_n$.}
\label{fig:Nusselt4}
\end{figure}

The secondary bifurcation at $\epsilon _n$ is strongly hysteretic. As
illustrated in figure~\ref{fig:Nusselt4}, when ${\epsilon}$ was
decreased from ${\epsilon} \ge {\epsilon}_n$, a transition back to
hexagons did not occur. Instead, the chaotic state persisted to values
of ${\epsilon}$ slightly below zero, at which point the conduction
state was reached (see also the solid triangles in
Fig.~\ref{fig:Bif_dia_detail}). In a separate section we will come
back to the hexagons.

\subsubsection{Determination of the Tricritical Point}

\begin{figure}
\epsfxsize = 3in
\centerline{\epsffile{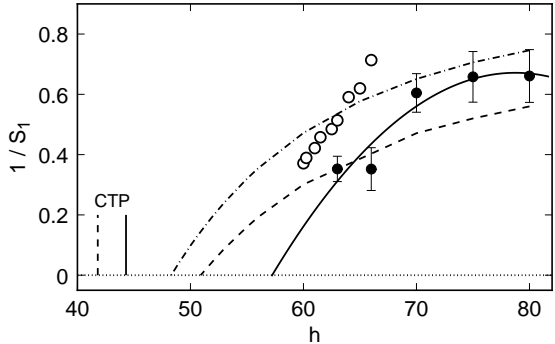}}  
\vskip 0.2in
\caption{The dependence on $h$ of the reciprocal of the initial slope $S_1$ of the Nusselt number as obtained from a fit of the data to Eq.~\protect{\ref{eq:nus_fit}}. The open circles are $1/S_{1,h}$ for cell 5, and the filled ones are $1/S_{1,r}$ for cell 6. For cell 5, data over the range $0 < \epsilon < \epsilon_n$ were used, and the pattern was hexagonal. For cell 6, data over the range $0.015 < \epsilon < \epsilon_n$ were used, and the pattern was one of rolls. The solid line represents a fit of a quadratic polynomial in $(h/h_t - 1)$ to the cell 6 $S_{1,r}$ data. The field $h_t = 57.2 \pm 2.6$, where $1/S_{1,r}$ for cell 6 extrapolates to zero, is interpreted to be the tricritical point. The theoretical results for $1/S_{1,r}$ are given by the dashed line. They yield a tricritical point at $h_t^{th} = 51$. The theoretical results for $1/S_{1,h}$ are given by the dash-dotted line. The location of the codimension-two point is given by the solid (experimental) and dashed (theoretical) short vertical lines. }
\label{fig:1/S_1}
\end{figure}

As $h$ approaches the tricritical point from above, the initial slope
$S_{1,r}$ of ${\cal N}$ for rolls is expected to diverge as $1/(h -
h_t)$. For cell 6, we estimated $S_{1,r}(h)$ from data for $\epsilon
\agt 0.015$ where rolls were observed.  At a given $h$, $S_{1,r}$ was
determined by fitting the polynomial

\begin{equation}
{\cal N} = 1 + S_{1,r} \epsilon + S_{2,r} \epsilon ^2 
\label{eq:nus_fit}
\end{equation}

\noindent with $\epsilon = \Delta T/\Delta T_c - 1$ to the data. The parameters $\Delta T_c$, $S_{1,r}$, and $S_{2,r}$ were adjusted
in the fit. Figure~\ref{fig:1/S_1} shows the dependence of $1/S_{1,r}$
upon $h$ as solid circles. The fitting procedure did not yield highly
accurate values because the Nusselt data for $\epsilon < 0.015$ had to
be excluded; thus the error bars for $S_{1,r}$ are relatively large.
The line is a fit of a quadratic polynomial in $(h/h_t - 1)$ to the results for
$1/S_{1,r}(h)$. This fit indicates the tricritical point to be at $h_t =
57.2 \pm 2.6$.

In order to compare the theory with the measurements above $h_t$, we calculated $S_{1,r}$ using the properties of 5CB. The results for $1/S_{1,r}$ are shown as a dashed line in Fig.~\ref{fig:1/S_1}. There is quite reasonable agreement with the experimental data for $h \agt 60$, particularly when it is considered that the initial slope of ${\cal N}$ in finite systems usually is smaller than the theoretical value for the infinite system. However, the theory yields a tricritical field $h_t^{th} = 51$ which differs significantly from the experimental estimates. We note that this difference is in the same direction as and somewhat larger than the corresponding one for the codimension-two point. We have no explanation for this difference.

\subsubsection {Hexagons}
In this section we will discuss in more detail the hexagonal patterns.
Hexagonal patterns at onset may be attributable to departures of the
physical system from the Oberbeck-Boussinesq (OB) approximation
\cite{OB,Bu67}, {\it i.e.} to a variations of the fluid properties over the
imposed temperature range.  For isotropic fluids it has been shown
that non-OB effects lead to hexagons at a transcritical (hysteretic)
primary bifurcation \cite{Bu67,Boden}. Below onset, for ${\epsilon}_a
\le {\epsilon} \le 0$, both hexagons and the conduction state are stable. Above
onset, hexagons are stable for $0 \le {\epsilon} \le {\epsilon}_r$.
For ${\epsilon}_r \le {\epsilon} \le {\epsilon}_b$ hexagons and rolls
are both stable, while for ${\epsilon} \ge {\epsilon}_b$ only rolls
are stable. When the thickness of the fluid layer is increased,
$\Delta T_c$ is reduced and thus departures from the OB approximation
become smaller. Thus the range of $\epsilon$ over which hexagons are
stable is reduced when the thickness of the fluid layer is increased,
as seen in the experiment by comparing cells 5 and 6.

A stability analysis of RBC with non-OB effects in a homeotropically
aligned NLC has not yet been carried out and would be very tedious.
Thus, in order to obtain at least a qualitative idea of the expected
range of stable hexagons, we used the theoretical results for the
isotropic fluid with the fluid properties of 5CB. The values of
${\epsilon}_a$, etc. are determined by a parameter ${\cal P}$ which
was defined by Busse \cite{Bu67} and is given by ${\cal P} = \Sigma_{i
  = 0}^4 \gamma_i {\cal P}_i \>,$ with $\gamma_0 = -\Delta \rho/\rho\ 
,$ $\gamma_1 = \Delta \alpha/2 \alpha\ ,$ $\gamma_2 = \Delta \nu/\nu\ 
,$ $\gamma_3 = \Delta \lambda/\lambda\ ,$ and $\gamma_4 = \Delta C_P /
C_P\ .$ Here $\rho$ is the density, $\alpha$ the thermal expansion
coefficient, $\nu$ the kinematic viscosity, $\lambda$ the
conductivity, and $C_P$ the heat capacity.  The quantities $\Delta
\rho$, {\it etc.} are the differences in the values of the properties
at the bottom (hot) and top (cold) end of the cell. For $\lambda$ we
used $\lambda_\parallel$, and for $\nu$ we used ${\alpha}_4 / 2 \rho$.
The coefficients ${\cal P}_i$ in the equation for $\cal P$ are given
by Busse. \cite{Bu67} However, here we use the more recent
results\cite{Tschammer} for large Prandtl numbers ${\cal P}_0 =
2.676$, ${\cal P}_1 = -6.631$, ${\cal P}_2 = 2.765$, ${\cal P}_3 =
9.540$, and ${\cal P}_4 = -6.225$ where ${\cal P}_3$ differs
significantly from the earlier calculation.

At the fields where the hexagons were observed, the temperature
difference across cell 5 (cell 6) was close to 5.05$^\circ$C
(2.07$^\circ$C). At these temperature differences, we obtained ${\cal
  P} = -1.5$, ${\epsilon}_a = -1.7\times 10^{-4}$, ${\epsilon}_r =
1.5\times 10^{-2}$, and ${\epsilon}_b = 5.3\times 10^{-2}$ for cell 5.
For cell 6 the values are ${\cal P} = -0.6$, ${\epsilon}_a =
-2.8\times 10^{-5}$, ${\epsilon}_r = 2.5\times 10^{-3}$, and
${\epsilon}_b = 8.7\times 10^{-3}$. From these estimates, it follows
that the hysteresis of size $\epsilon_a$ is too small to be noticeable
in either cell with our resolution. The largest value of $\epsilon$ at
which hexagons could exist in cell 6 would be $\epsilon = 8.7\times
10^{-3}$. However, we observe hexagons to exist to nearly twice this
value. If the same is true for the thinner cell, the hexagon-roll
transition attributable to non-OB effects should happen at a value of
$\epsilon$ greater than $\epsilon_n(h)$ for the field range over which
the experiments were performed. Thus, instead of leading to a time
independent state, as observed in cell 6, the hexagon-roll transition
is preceded by a transition to a state exhibiting spatio-temporal
chaos at $\epsilon_n$.

The hexagonal pattern may be regarded as a superposition of three sets
of rolls with amplitudes $A_i$, $i = 1, 2, 3$, corresponding to the
three basis vectors at angles of $120^\circ$ to each other.
According to the Landau model\cite{CH92} the steady-state amplitudes $A_i$ are determined by
\begin{equation}
\epsilon A_1 + {1 \over 2}b(A_2^2 +A_3^2) -g_{11}A_1^3 - g_{12}A_1 A_2^2 -g_{13}A_1A_3^2 = 0 
\end{equation}
and the corresponding cyclic permutation for $i =2,3$.
Since all amplitudes are expected to be equal in hexagons 
($A_1 =A_2 = A_3 =A$), one has
\begin{equation}
\epsilon A + b A^2 - (g + 2 \tilde g)A^3 = 0 
\end{equation}
where $g$ is the
self-coupling coefficient $g_{11}$ and $\tilde g$ is the cross-coupling
coefficient $g_{12} = g_{13}$. Because of the term $b A^2$, the
bifurcation is transcritical and thus hysteretic. However, as we
discussed above and as is shown for instance by the data in
Fig.~\ref{fig:Supercrit}, this effect is not resolved in the
experiment because the coefficient $b$ (which is determined by ${\cal
  P}$) is too small. Thus we neglect the term $b A^2$, and have to a
good approximation
\begin{equation}
{\cal N} - 1 =3 A^2 = {{3 \epsilon} \over {g + 2 \tilde g}} .$$
\end{equation}

At the tricritical point $g$ vanishes as $g = g_0(h - h_t)$. However,
there is no reason why $\tilde g$ should vanish also at $h_t$. Thus,
one would expect the slope $S_{1,h} = 3 / (g + 2 \tilde g)$ of ${\cal N}$ near $\epsilon = 0$
to remain finite at $h_t$ and equal to $3 / 2 \tilde g$.
To test this idea, we fitted ${\cal N}$ for cell 5 over the $\epsilon$
range where hexagons were observed (i.e. up to $\epsilon_n$) to an
equation like Eq.~\ref{eq:nus_fit}. Since data quite close to
threshold could be used, the results for $S_{1,h}$ are much more
precise than those for cell 6. They are given in Fig.~\ref{fig:1/S_1}
as open circles. One can see that $1/S_{1,h}$ is non-zero at $h_t$. It
extrapolates to zero near $h = 53$, which is well below $h_t = 57.2
\pm 2.6$. Unfortunately the relatively large uncertainty of $h_t$ and
$S_{1,r}$ prevents the accurate determination of $\tilde g$. At $h_t$,
we find $\tilde g = 3/2 S_{1,h} \simeq 0.3$. With increasing $h$,
$\tilde g$ also increases. For instance, the data in
Fig.~\ref{fig:1/S_1} suggest that $\tilde g = 3 /2 S_{1,h} - 1/2
S_{1,r} \simeq 0.8$ for $h = 66$.

The experimental results for $1/S_{1,h}$ cannot agree quantitatively with the theory because we already know that  $h_t^{th}$ is
lower than the experimental value. Nevertheless we calculated $1/S_{1,h}$, and give it in Fig.~\ref{fig:1/S_1} as the dash-dotted line. We see that the relationship between $1/S_{1,h}$ and $1/S_{1,r}$ is quite similar in theory and experiment.

\subsection{The High-Field Limit of $R_c$ and $k_c$}

It is highly probable that there exists a high-field regime where the convection
phenomena become independent of the field since the director
is then frozen in the homeotropic configuration. 
It is instructive to examine whether the experimental data extend to sufficiently high fields to fully reveal this behavior. Building on the results of FDPK \cite{FDPK92}, one
can show that in a one-mode approximation the neutral curve in the
limit $h \rightarrow \infty$ is given by

\begin{eqnarray}
&R&_{c,\infty}(k) = {{(\lambda_\perp/\lambda_\parallel)k^2+\pi^2}\over{2k^2 I_1^2}}\nonumber\\ 
&\times& \left [-I_2 k^2 \left ({2 \eta_1 \over \alpha_4} + {2 \eta_2 \over \alpha_4} + {2 \alpha_1 \over \alpha_4}\right ) + k^4 {2 \eta_2 \over \alpha_4} + b^4 {2 \eta_1 \over \alpha_4}\right ].
\label{eq:Rcinf}
\end{eqnarray}

\noindent Here $\eta_1 = (\alpha_4 + \alpha_5 - \alpha_2)/2$ and  $\eta_2 = (\alpha_3
+ \alpha_4 + \alpha_6)/2$ are Miesowicz viscosity coefficients,\cite{Ge73} and   $I_1 = 0.69738, I_2 = -12.3026$, and
$b = 1.5056\pi$.  This leads to $R_{c,\infty}(k_c) = 3090.6$ and
$k_{c,\infty} = 4.294$. Using many modes, one obtains numerically
$R_{c,\infty}(k_c) = 3056.6$ and $k_{c,\infty} = 4.328$, close to the
one-mode result. One can see that only the viscosities enter into
$R_{c,\infty}$, and not the elastic constants. This is so because the
director is held rigid by the field. $R_{c,\infty}$ is larger than the
isotropic-fluid value because of the additional viscous interaction
between the flow and the rigid director field.

In the high-field limit, we obtain 
\begin{equation}
R_c = R_{c,\infty} + R_1/h^2 + {\cal O}(h^{-4})\ . 
\end{equation}
The coefficient $R_1$ has not been calculated in detail, but is proportional to $k_{33}^{-1}$. The fact that at order $1/h^2$ elastic constants enter suggests the beginning of some director distortion by the flow. 

\epsfxsize = 3in
\begin{figure}
\centerline{\epsffile{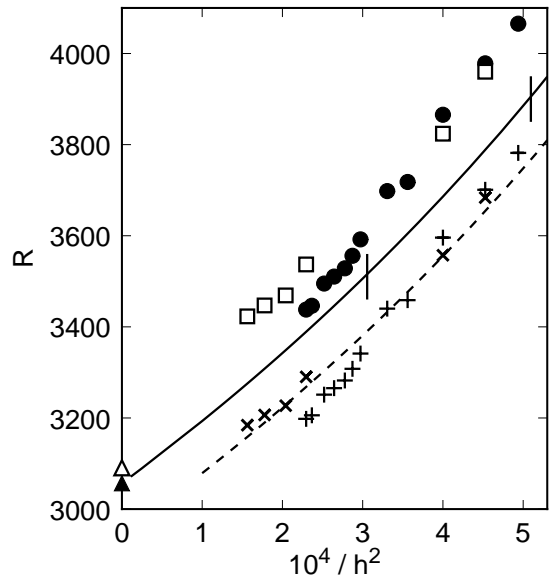}}  
\caption{The critical Rayleigh number as a function of $1/h^2$. The solid line was calculated using the parameters of Ref. \protect{\cite{FN1}}. The open triangle is $R_{c,\infty}$ from Eq.~\protect{\ref{eq:Rcinf}}. The solid triangle is the many-mode numerical result for $R_{c,\infty}$. The solid circles are from cell 5, and the open squares are from cell 6. The vertical bars indicate the location of the codimension-two point (right bar) and the tricritical point (left bar). The plusses, crosses, and dashed line show what happens to the data and the theoretical curve if the viscosity $\alpha_4$ is increased by 7.5\%.}
\label{fig:Rc_vs_h-2}
\end{figure}

\begin{figure}
\epsfxsize = 3in
\centerline{\epsffile{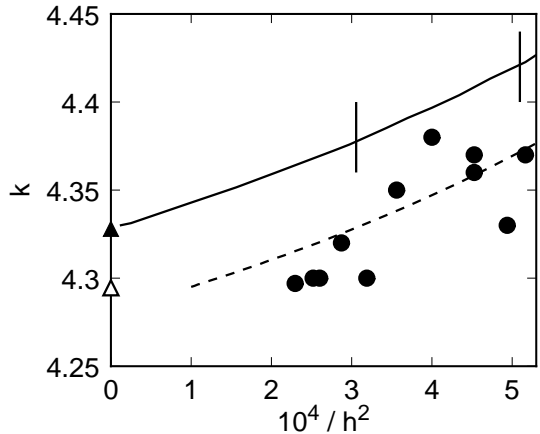}}  
\vskip 0.2in
\caption{The critical wavenumber as a function of $1/h^2$. The solid line was calculated using the parameters of Ref. \protect{\cite{FN1}}. The open triangle is $k_{c,\infty}$ from Eq.~\protect{\ref{eq:Rcinf}}. The solid triangle is the many-mode numerical result for $k_{c,\infty}$. The vertical bars indicate the location of the codimension-two point (right bar) and the tricritical point (left bar). The dashed line shows what happens to the theoretical curve if the viscosity $\alpha_4$ is increased by 7.5\% (the data for $k_c$ remain unchanged).}
\label{fig:kc_vs_h-2}
\end{figure}

In Figs.~\ref{fig:Rc_vs_h-2} and \ref{fig:kc_vs_h-2} we show the
experimental data and theoretical results as a function of $h^{-2}$.
The one-mode high-field limits $R_{c,\infty}$ and $k_{c,\infty}$ are
shown in the figures as open triangles. The corresponding numerical
many-mode results are given as solid triangles. The experimental data
for $R_c$ and $k_c$ are consistent with the expected dependence on $h$,
but respectively fall about 4\% above and 1\% below the calculation. The data for
$R_c$ and $k_c$ at the highest experimental field $h \simeq 80$ are
already within about 6\% and 1\% respectively of the infinite-field
value. Thus it seems unlikely that qualitatively new phenomena could
be discovered by measurements at even higher fields.

We examined whether the small difference between the theory and the
experiment could be removed by small adjustments in the values of the
fluid properties. We found that an increase by 7.5\% of $\alpha_4$
yielded the plusses and crosses for the data in
Fig.~\ref{fig:Rc_vs_h-2} and the dashed lines in
Figs.~\ref{fig:Rc_vs_h-2} and \ref{fig:kc_vs_h-2} (the data for $k_c$
in Fig.~\ref{fig:kc_vs_h-2} are not affected by changing $\alpha_4$).
The adjustment of $\alpha_4$ produced an excellent fit for both $k_c$
and $R_c$. However, it spoiled the excellent agreement for $R_c$ along
the oscillatory branch below the codimension-two point shown in
Fig.~\ref{fig:Bif_dia} and did not significantly reduce the difference
between calculation and experiment for $k_c$ at small $h$ which is
shown in Fig.~\ref{fig:K_v_h}. Various other attempts to adjust the fluid properties used in the theoretical calculations were unsuccessful 
in yielding improved overall agreement between theory and experiment 
over the entire field range.  

\subsection{Nonlinear States at High Fields}

\begin{figure}
\epsfxsize = 3in
\centerline{\epsffile{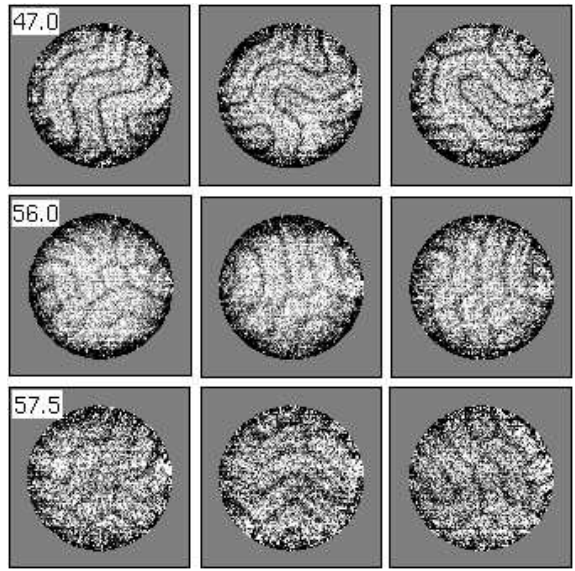}}  
\vskip 0.2in
\caption{Time sequences of the spatially and temporally chaotic flow for $R = 4000$ and at different values of $h$ for cell 5. The images in each row were taken at the field indicated in the leftmost image. Time increases from left to right. Images were taken in one hour intervals. The wavenumbers of the patterns are $h = 47$: $k_n = 3.4$, $h = 56$: $k_n =3.6$, $h = 57.5$: $k_n =3.9$.}
\label{fig:STC}
\end{figure}

Beyond the codimension-two point the finite-amplitude flow was split into three regions distinguished by their characteristic wavenumbers $k_n \simeq 3.4, 3.6$ and 3.9. These regions are shown in Fig.~\ref{fig:Bif_dia_detail}.  Figure~\ref{fig:STC} shows some characteristic patterns. Qualitatively the patterns appear similar, each exhibiting spatio-temporal chaos.
The transitions between these state depended on both $h$ and $R$. They were determined by measuring the change in $k_n$ as $h$ was varied at a given Rayleigh number. As already mentioned above, to our knowledge there are no theoretical predictions for these patterns.

\section{Acknowledgment}

One of us (WP) acknowledges support from a NATO Collaborative Research Grant. One of us (GA) is grateful to the Alexander von Humboldt Foundation for support. The work in Santa Barbara was supported by the National Science Foundation through Grant DMR94-19168.

\end{multicols}


\begin{references}
\bibitem{FN1}The properties were evaluated from the formulas in Appendix B of Ref. \cite{Ah96} for $T = 25.6^\circ$C and $T_{NI} = 35.17^\circ$C. For the elastic constants we have $(k_{11},k_{22},k_{33}) = (0.609,0.403,0.880)~\mu$dynes. For the Lesley viscosity coefficients we used $(\alpha_1, \alpha_2, \alpha_3, \alpha_4, \alpha_5, \alpha_6) =$  (-0.0718, -0.804,\break -0.0426, 0.651, 0.651, -0.19) dyne s/cm$^2$. The density was 1.022 g/cm$^3$, and the expansion coefficient was $\alpha = 0.895\times 10^{-3}$ K$^{-1}$. We used $(\kappa_\perp,\kappa_\parallel) = (0.677\times 10^{-3},1.136\times 10^{-3})$ cm$^2$/s and $\chi_a = 1.132\times 10^{-7}$ cm$^3$/g.
\bibitem{FN}Cell 4 was used to obtain the results reported in Ref. \cite{Ah96}, although since that work it had been disassembled and the sample had been changed.
\bibitem{FN3} For small $\epsilon > 0$ the unstable pure conduction state was extremely long lived because it had to evolve from spontaneous fluctuations. For $\epsilon \alt 0.015$ its lifetime exceeded the
five-hour equilibration time used in the experiment, and thus there are no data with increasing $\Delta T$ in this $\epsilon$-range. With decreasing $\Delta T$, the decay of the amplitude to its new steady state was relatively fast and data very near $\epsilon = 0$ could be taken.
\end{references}
\end{document}